\documentclass[%
 reprint,
 amsmath,amssymb,
 aps,
 prb,
]{revtex4-2}

\usepackage{graphicx}
\usepackage{dcolumn}
\usepackage{bm}
\usepackage{hyperref}

\begin{document}


\title{Exploring terahertz-scale exchange resonances in synthetic ferrimagnets with ultrashort optically induced spin currents}

\author{Julian Hintermayr}
\email[Electronic mail: ]{j.hintermayr@tue.nl}
\affiliation{%
 Department of Applied Physics, Eindhoven University of Technology,\\
 P.O. Box 13, 5600 MB Eindhoven, the Netherlands
}%
\author{Youri L.\,W. van Hees}%
\affiliation{%
 Department of Applied Physics, Eindhoven University of Technology,\\
 P.O. Box 13, 5600 MB Eindhoven, the Netherlands
}%
\author{Bert Koopmans}
\affiliation{%
 Department of Applied Physics, Eindhoven University of Technology,\\
 P.O. Box 13, 5600 MB Eindhoven, the Netherlands
}%

\date{\today}

\begin{abstract}
Using spin currents generated by fs laser pulses, we demonstrate excitation of GHz ferromagnetic resonance and THz ferrimagnetic exchange resonances in Co/Gd/Co/Gd multilayers by time-resolved magneto-optic Kerr effect measurements.
Varying the Gd layer thickness allows for a tuning of the resonance spectrum by manipulating the total angular momentum and strength of effective exchange fields between the antiferromagnetically coupled layers.
Close to the compensation point of angular momentum, a minimum in the frequency of the exchange-dominated mode and a maximum in the frequency of the ferromagnetic resonance mode is observed.
Finally, to gain better understanding of the excitation mechanism, we analyze the anomalous variation in the measured exchange mode amplitude as a function of its frequency. A peak in this amplitude in the vicinity of the compensation point of angular momentum is explained using a macrospin model, taking nonlinear effects at finite precession amplitudes into account.  
\end{abstract}

\maketitle

\section{Introduction}
Ferrimagnets combine a number of advantages of ferro- and antiferromagnets, making them a promising platform for fast and easily integratable spintronic devices~\cite{Finley:2020, Kim:2022, Zhang:2023}. Strong, alternating exchange fields are inherent to antiferromagnets, giving rise to high-frequency ($\sim$THz) exchange resonance modes (EXMs)~\cite{Gomonay:2014, Keffer:1952}. Their lack of net spin polarization however makes experimental access challenging~\cite{Jungwirth:2016}. Metallic \textit{ferrimagnets}, in contrast, can exhibit finite conduction spin polarization --- even at the compensation point of angular momentum --- and can thus easily be probed by magneto-optic or magneto-resistive effects, given that they consist of sublattices with different atomic species~\cite{Finley:2020}.

While the high frequency makes this class of materials a promising candidate for the development of THz spintronic devices, techniques of exciting resonances at such high frequencies coherently are sparse. In the past, EXMs have been excited for instance by thermal laser excitation in combination with an applied field~\cite{Mekonnen:2011, Stanciu:2006} and circularly polarized laser pulses~\cite{Kimel:2005, Deb:2016, Parchenko:2016}. The former approach only made it possible to excite EXMs in the $<100$~GHz range, whereas the latter approach requires materials with significant inverse Faraday effect. In addition, pulses of intense THz radiation have been shown to facilitate excitation of fast spin dynamics~\cite{Kampfrath:2011, Baierl:2016, Blank:2021}.

In this work, we will make use of optically induced spin currents that are generated by a neighboring ferromagnet upon ultrafast demagnetization~\cite{Malinowski:2008, Kampfrath:2013, Choi:2014, Choi:2015}. The optically induced spin-transfer torques (OSTTs) generated in this manner have been used to excite exchange-dominated standing spin waves with frequencies exceeding 1~THz~\cite{Lalieu:2017b, Lalieu:2019, Brandt:2021}. Moreover, its has been shown that such spin currents can be used to assist in single-shot all-optical switching of ferrimagnets~\cite{vanHees:2020,Remy:2020} or even induce full switching in ferromagnets~\cite{Remy:2020, Igarashi:2020, Remy:2023, Igarashi:2023}.
As we will show, a synthetic ferrimagnetic quadlayer consisting of the transition metal (TM) Co and the rare earth (RE) Gd can host not only GHz ferromagnetic resonance but also THz exchange resonance modes, in spite of the arguably weaker exchange coupling compared to CoGd alloys. Co/Gd multilayers are highly relevant for a variety of spintronic applications as they exhibit energy-efficient single-shot all-optical magnetization switching~\cite{Lalieu:2017}. Additionally, ultrafast domain wall motion close to the compensation point of angular momentum has been observed very recently~\cite{Li:2022}. The coexistence of these properties makes this material a promising candidate for hybrid spintronic-photonic memory applications~\cite{Lalieu:2019b, Li:2022}.

In the following, we will show that OSTTs are an excellent tool to excite and study GHz ferromagnetic resonance (FMR) and THz EXMs in Co/Gd multilayers. Furthermore, we explain our findings with an analytical macrospin model.

\section{Sample structure and characterization}
\begin{figure*}[htbp]
    \centering
    \includegraphics[width=17.0cm]{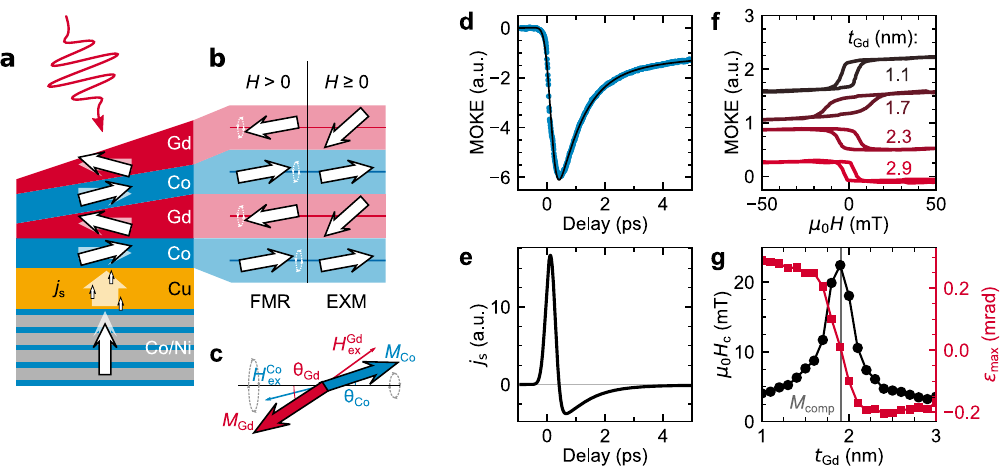}
    \caption{Cartoon of \textbf{a},~sample architecture showing local arrangements of magnetization and the resonance excitation mechanism \textbf{b},~stable spin precession configurations and \textbf{c},~local directions of exchange fields in canted magnetization states between Co and Gd, allowing for fast resonance frequencies. \textbf{d},~Demagnetization trace of the [Co/Ni] injection layer, based on which the optically induced spin current pulse in \textbf{e} is calculated. \textbf{f},~In-plane hysteresis loops of the Co/Gd/Co/Gd absorption layer for different Gd thicknesses measured with L-MOKE. \textbf{g},~Coercive field and MOKE step as a function of Gd thickness. The vertical gray line indicates the magnetization compensation point $M_\mathrm{comp}$.}
    \label{fig:sample}
\end{figure*}

The idea behind the sample design is to layer a ferrimagnet with an in-plane (IP) magnetic easy axis onto a ferromagnet with perpendicular magnetic anisotropy (PMA) that efficiently generates spin currents upon ultrafast demagnetization~\cite{Schellekens:2014b, Choi:2014}. We chose a synthetic ferrimagnetic \textit{quadlayer} of Co/Gd/Co/Gd as it offers the possibility of reaching magnetic compensation by fine-tuning individual layer thicknesses while maintaining IP anisotropy. The spin current generation layer is a Pt/[Co/Ni]$_4$ multilayer which provides PMA at relatively large magnetic volumes~\cite{Lalieu:2017b}.
Between the two layers, a Cu spacer is placed to magnetically decouple the two systems while allowing for spin currents to pass. The final material stack used in this study is as follows: Si:B/Ta(4)/Pt(4)/[Co(0.2)/Ni(0.6)]$_4$/Co(0.2)/Cu(5)/ [Co(0.8)/Gd($t_\mathrm{Gd}$)]$_2$/TaN$_x$(5) (thicknesses in nm) with Gd thicknesses ranging from 0--3~nm (schematically shown in Fig.~\ref{fig:sample}~\textbf{a}). Layers are deposited by dc magnetron sputter deposition at room temperature. The demagnetization is triggered and measured by $\sim 100$~fs laser pulses with a central wavelength of 780~nm and 80~MHz repetition rate, using a standard pump-probe setup to measure the time-resolved magneto-optic Kerr effect (TR-MOKE). As the MOKE of Gd at this wavelength is small~\cite{Erskine:1973}, we mostly probe the signal arising from Co and Co/Ni layers. Pump and probe pulses are focused on the sample surface with near normal incidence.

Firstly, we investigate the demagnetization trace and derive the optically induced spin current profile to understand the excitation mechanism. The resulting trace is shown in Fig~\ref{fig:sample}~\textbf{d}, revealing rapid demagnetization within the first 300~fs and a slower remagnetization over several ps. We assume that the spin current originating from this process follows the spin-pumping model, also referred to as ``$\mathrm{d}M/\mathrm{d}t$'' model~\cite{Choi:2014, Choi:2015, Lichtenberg:2022, Beens:2022}, where the demagnetization is explained by excitation of magnons that transfer their angular momentum to mobile conduction electrons. The shape of such a spin current pulse is thus derived from the time derivative of the demagnetization curve (Fig.~\ref{fig:sample}~\textbf{a}) and shown in Fig.~\ref{fig:sample}~\textbf{e}. De- and remagnetization give rise to a positive and negative peak in spin current respectively, resulting in a bipolar pulse on the ps timescale. It is thus suited to excite THz-scale dynamics. To characterize the static magnetic properties of the Co/Gd absorption stack, MOKE hysteresis loops in longitudinal geometry are recorded for different Gd thicknesses. Measurements are shown in Fig.~\ref{fig:sample}~\textbf{f}. We find that the easy axis of the absorption stack lies in-plane (IP), with a magnetic compensation point ($M_\mathrm{comp}$) at around 1.9~nm Gd. At this thickness, the magnetic moments of the antiferromagnetically coupled Co and Gd layers exactly cancel each other out. The switch in sign of the measured Kerr ellipticity $\varepsilon_\mathrm{max}$ is accompanied by a maximum in coercive field, as shown in Fig.~\ref{fig:sample}~\textbf{g}, where $M_\mathrm{comp}$ is indicated by a vertical line. The angular momentum compensation point $L_\mathrm{comp}$ is expected to lie at slightly lower Gd thicknesses, as $g_\mathrm{Co}>g_\mathrm{Gd}$ (see Appendix). Due to the fact that the exact magnetization profile in the stack is not known, it is not possible to precisely determine $L_\mathrm{comp}$. We note that, even though the samples are deposited as multilayers, significant intermixing between Co and Gd is expected at the interface when sputter depositing at room temperature~\cite{Colino:1999, Pelka:2001, Andres:2002, Singh:2019}. As the Co thickness is only 0.8~nm, the intermixing depth in thin RE/TM multilayers is expected to be in a similar order of magnitude as the individual layer thickness, leading to significant alloy-like regions~\cite{Kools:2022}.

\section{Spin resonance modes}
Having discussed the static magnetic properties of the synthetic ferrimagnetic multilayer, we will introduce the types of spin resonances that can occur in this system. Generally, two types of precessions are possible within a two-sublattice spin system with antiferromagnetic coupling. For simplicity, we limit ourselves to the alloy-like case for now, assuming one macrospin per material, and neglect magnetic anisotropy. Further discussion on the validity of this approximation is provided later in this work. In the presence of an external magnetic field, FMR-like dynamics are allowed, where the magnetization of Co and Gd sublattices, $M_\mathrm{Co}$ and $M_\mathrm{Gd}$, are aligned fully antiparallel and precess around the direction of the external field $H$ (schematically shown in Fig.~\ref{fig:sample}~\textbf{b}). The frequency of this oscillation can be approximated as~\cite{Gurevich:1996}
\begin{equation}\label{eq:FMR}
    f_\mathrm{FMR}=\frac{\mu_0}{2\pi}\,\frac{M_\mathrm{Co} - M_\mathrm{Gd}}{\frac{M_\mathrm{Co}}{\gamma_\mathrm{Co}} -  \frac{M_\mathrm{Gd}}{\gamma_\mathrm{Gd}}}\,H = \frac{\mu_0}{2\pi}\,\gamma_\mathrm{eff}\,H.
\end{equation}
$\mu_0$ denotes the vacuum permeability, $\gamma_\mathrm{Co, Gd}$ the gyromagnetic ratios of Co and Gd, and $\gamma_\mathrm{eff}$ the effective gyromagnetic ratio. Given that $\gamma_\mathrm{Co}\neq\gamma_\mathrm{Gd}$, the magnetization and angular momentum of the two sublattices compensate at different atomic fractions. As we approach the point where the total angular momentum is fully compensated ($L_\mathrm{comp}$), the torque which is proportional to $\mathbf{M}_\mathrm{eff}\times \mathbf{H}$ remains finite, resulting in a divergence in $\gamma_\mathrm{eff}$, as predicted by eq.~(\ref{eq:FMR}). Please note that this approximation neglects demagnetizing fields and becomes invalid very close to $L_\mathrm{comp}$. A more involved treatment removes the divergence and leads to only a pronounced maximum in FMR frequency at $L_\mathrm{comp}$~\cite{Schlickeiser:2012}.

The exchange-dominated resonance mode with typically much higher frequencies requires $M_\mathrm{Co}$ to be at an angle with the exchange field it experiences. This field is proportional to the Weiss constant $\lambda$, indicating sign and strength of the exchange interaction, and can be expressed as $\mathbf{H}_\mathrm{ex}^\mathrm{Gd}=\lambda \mathbf{M}_\mathrm{Gd}$. Figure~\ref{fig:sample}~\textbf{b} and \textbf{c} schematically show the canted alignments of spins and exchange fields that are necessary for this mode to exist. Its frequency is given by~\cite{Gurevich:1996}
\begin{equation}\label{eq:EXM}
    f_\mathrm{EXM}=\frac{\mu_0\lambda\gamma_\mathrm{Co}\gamma_\mathrm{Gd}}{2\pi}\left(\frac{M_\mathrm{Co}}{\gamma_\mathrm{Co}} -  \frac{M_\mathrm{Gd}}{\gamma_\mathrm{Gd}}\right)
\end{equation}
in the limit of small oscillation angles. In contrast to the FMR mode, the EXM frequency vanishes at the angular momentum compensation point. The change in sign in frequency predicted by the dispersion relation indicates an intriguing change in handedness of the oscillation, which has been experimentally verified by Brillouin light scattering~\cite{Kim:2020, Haltz:2022}.

The effective damping parameter $\alpha_\mathrm{eff}$ in this system has been calculated to show a maximum at $L_\mathrm{comp}$ for both FMR and EXM modes~\cite{Schlickeiser:2012}. Spin resonances are therefore expected to only be very short-lived close to compensation. If the individual sublattice damping coefficients are identical, that is, $\alpha_\mathrm{Co}=\alpha_\mathrm{Gd}$, $\alpha_\mathrm{eff}$ of FMR and EXM will likewise be equal. If they differ from one another, the ratio of damping parameters of FMR and EXM will depend on $\alpha_\mathrm{Co}/\alpha_\mathrm{Gd}$.

We note that the symmetry of the quadlayer system allows in principle for higher-order exchange resonances where, for instance, the two Co layers are precessing out of phase. However, the frequencies of such modes are expected to be far above those of the fundamental EXM. We therefore cannot excite them with our technique, assuming the wedged Gd does not become thick enough to only provide a weak link between the Co layers.
Yet another type of exchange-driven modes that can exist in thin films are quantum-confined standing spin waves that have been excited and measured with techniques similar to those used in this work~\cite{Lalieu:2017b, Lalieu:2019, Brandt:2021}, but for a 0.8 nm Co layer expected frequencies ($>10$~THz) are too high to be excited by our method. Also, more complex modes delocalized throughout the whole ferrimagnetic stack have not been identified in our experiments, and we find a description in terms of an alloy with only two sublattices sufficient to explain all our results.
%

\section{Results}

\subsection{Time-resolved magnetization dynamics}

To investigate the dynamics of the sample, we again employ pump-probe spectroscopy. As mentioned previously, resonances are excited by injection of spin current pulses generated in the Co/Ni layer into the Co/Gd absorption layer. The current from the injection layer shows OOP spin polarization and exerts an OSTT on the IP absorption layer, canting the magnetization OOP away from its ground state, as illustrated in Fig.~\ref{fig:sample}~\textbf{a}. If the effective magnetization and the applied field are at an angle in the excited state, the FMR mode is triggered. If $M_\mathrm{Co}$ and $M_\mathrm{Gd}$ are at an angle, the following precession is described by the EXM. We note that the ultrafast laser excitation of the Co/Gd quadlayer could lead to an injection of an IP-polarized spin current into the Co/Ni layer as well. Since the dynamics in the Co/Ni layer induced by the following OSTT are expected to be fully IP as well, the dynamics are not probed during the experiment due to the polar measurement geometry.

Before the measurement, we saturate the injection stack in the positive z-direction and apply an IP field for FMR measurements and no IP field for EXM measurements. We then repeat the measurement with the injection stack saturated in the opposite direction and calculate the magnetic signal as the difference of measurements with positively and negatively saturated injection layer. The sum of the measurements includes non-magnetic artefacts such as the coherence peak that occurs during pump-probe overlap~\cite{Eichler:1984, Luo:2009} and is disregarded.

\begin{figure}[htbp]
    \centering
    \includegraphics[width=8.6cm]{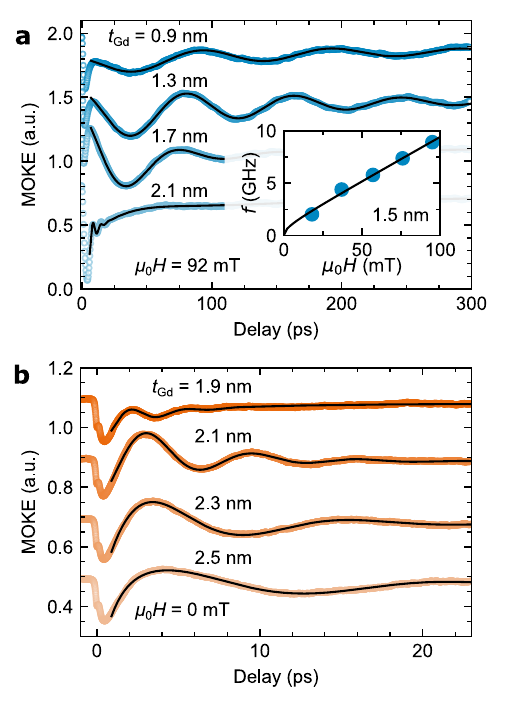}
    \caption{\textbf{a}~Time-resolved MOKE measurements of homogeneous FMR precessions in the in-plane Co/Gd/Co/Gd layer at a magnetic field of 92~mT. Offsets are proportional to the Gd thickness. The inset shows the precession frequency as a function of applied field at a fixed Gd thickness of 1.5~nm including a fit according to eq.~(\ref{eq:Kitt}). \textbf{b}~Oscillation measurements of exchange modes using complex MOKE. Black lines represent damped sine fits. Note the different timescales in \textbf{a} and \textbf{b}.}
    \label{fig:FMR}
\end{figure}

\begin{figure}[htbp]
    \centering
    \includegraphics[width=7.0cm]{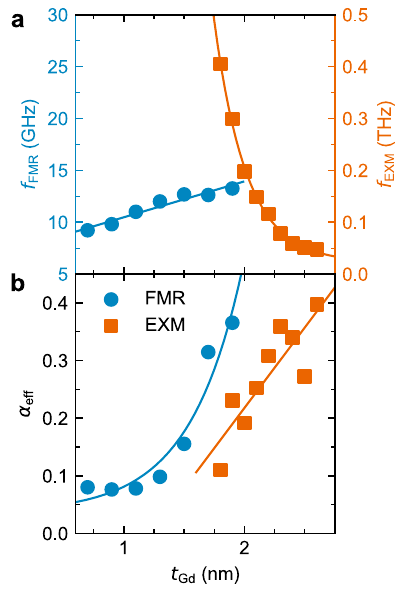}
    \caption{\textbf{a}~Frequency of the ferromagnetic resonance mode in a field of 92~mT and that of the exchange mode in the absence of a field as a function of Gd thickness. Please note the different y-scales. \textbf{b}~Effective damping parameter of said modes. Lines are guides for the eye.}
    \label{fig:freq}
\end{figure}

Homogeneous, FMR-like precession modes with an applied IP field of 92~mT are shown in Fig.~\ref{fig:FMR}~\textbf{a} for a range of Gd thicknesses. During the first few ps, the demagnetization of the injection layer dominates the signal. Subsequently, the oscillations of the absorption layer are observed. For a more detailed analysis of the measurements, we fit functions of the form
\begin{equation}
    A\sin (2\pi ft - \varphi)\mathrm{e}^{-t/\tau}+B\,\mathrm{e}^{-t/C}+D
\end{equation}
to our data (black lines), allowing an extraction of the frequency $f$ and the effective damping parameter $\alpha_\mathrm{eff}=1/2\pi f\tau$. $A$ denotes the oscillation amplitude, $\varphi$ the phase, and $\tau$ the decay time of the oscillation. The following terms capture the remagnetization behavior of the injection layer with an amplitude of $B$, a decay time of $C$ and a constant offset $D$. A steady increase of the FMR frequency as a function of $t_\mathrm{Gd}$ is observed, which is explained by an increase in Gd angular momentum, whereby the total angular momentum decreases. In accordance to eq.~(\ref{eq:FMR}) $\gamma_\mathrm{eff}$ increases, and with it the frequency and effective damping, as observed in Fig.~\ref{fig:freq}~\textbf{a} and~\textbf{b}.

For a Gd thickness of 1.5~nm, FMR measurements at different fields were recorded. The extracted frequencies as a function of applied field are plotted in the inset of Fig.~\ref{fig:FMR}~\textbf{a}. As a linear fit according to eq.~(\ref{eq:FMR}) yielded unsatisfactory results due to neglecting demagnetizing fields, we used the following Kittel-like equation, accurately implementing the effect of finite demagnetizing effects in thin films, where the effective magnetization $M_\mathrm{eff}$ and $\gamma_\mathrm{eff}$ are fitted:
\begin{equation}\label{eq:Kitt}
    f_\mathrm{Kitt}=\frac{\mu_0\gamma_\mathrm{eff}}{2\pi}\sqrt{H(H+M_\mathrm{eff})}.
\end{equation} 
The best fit is found for $\gamma_\mathrm{eff}/\gamma_\mathrm{e}=3.0$, with the electron gyromagnetic ratio $\gamma_\mathrm{e}$. This strong boost of $\gamma_\mathrm{eff}$ is expected as $L_\mathrm{comp}$ is being approached. Furthermore, we extract $M_\mathrm{eff}=25~\mathrm{kA/m}$, which corresponds to $\sim 2\%$ of the saturation magnetization of elemental Co. Additional possible anisotropy-related effects arising from the multilayered nature of the sample are also captured in $M_\mathrm{eff}$. They may stem for instance from interfacial anisotropy which could give both positive or negative contributions to $M_\mathrm{eff}$ as well as shape anisotropy. The very low value of $M_\mathrm{eff}$ implies, however, that the sample is close to $M_\mathrm{comp}$ and that the anisotropy contributions are small.

Upon increasing the Gd thickness to 2.1~nm, a drastic discontinuous change in the oscillation frequency is observed (see Fig~\ref{fig:FMR}~\textbf{a}). We associate this observation with the onset of the EXM and the simultaneous suppression or over-damping of the FMR mode. To increase our sensitivity to the EXM at short time delays, a quarter-wave plate in the probe beam path is used to mostly cancel out the contribution of the demagnetization of the injection layer to the signal. This technique is known as complex MOKE, for further information the reader is referred to Refs.~\cite{Hamrle:2002, Schellekens:2014}. Figure~\ref{fig:FMR}~\textbf{b} shows oscillation measurements of EXM resonances acquired in this manner. During the first ps, some unavoidable leaking signal from the injection layer and remanent artefacts from the coherence peak close to zero delay are visible in the signal. Thereafter, damped oscillations are observed. Please note the different scale of the x-axis compared to Subfigure~\textbf{a}. Increasing Gd thicknesses results in a strong decrease of precession frequency and a variation of the precession amplitude. The decrease of frequency is well explained by an approach of the angular momentum compensation point as predicted by eq.~(\ref{eq:EXM}). Plotting the resonance frequencies as a function of $t_\mathrm{Gd}$ (see Fig.~\ref{fig:freq}~\textbf{b}) shows a highly non-linear behaviour. Two factors give rise to such nonlinearity: Firstly, we recall that the Curie temperature of Gd is below room temperature. Thus, it only shows a finite magnetization within the partially intermixed Co/Gd and Gd/Co interfaces as well as proximity-induced magnetization in the regions close to the interfaces. As one moves away, this induced magnetization---and thus the angular momentum---decreases exponentially~\cite{Kools:2022}. Secondly, increasing the Gd thickness decreases the average exchange parameter $\lambda$ in eq.~(\ref{eq:EXM}) which scales inversely proportional to the thickness in magnetic multilayers. Please note that this coupling parameter only influences the EXM and not the FMR frequency. Hence, this effect does not influence the thickness dependence of the FMR mode.

Another peculiarity of the EXM frequency is that it decreases monotonously, implying either that the angular momentum compensation point is being approached, yet never crossed, or that oscillation amplitudes are too large for the small angle approximation to hold that predicts a vanishing frequency. The magnetostatic characterization, on the other hand, clearly revealed a magnetic compensation point  around 1.9~nm Gd. Since $\gamma_\mathrm{Co}>\gamma_\mathrm{Gd}$, $L_\mathrm{comp}$ should lie at even smaller $t_\mathrm{Gd}$ than $M_\mathrm{comp}$. However, the magnetization of Gd is strongly temperature-dependent, especially in TM/Gd multilayers. Shortly after the arrival of the laser pulse it is therefore very well possible that the system is Co dominant even though the room temperature characterization revealed Gd-dominance. Thus, the absence of an angular momentum compensation point in time-resolved measurements at strongly increased temperatures during the first ps after laser irradiation could result in a Co-dominated sample across all investigated Gd thicknesses. A dependence of the frequency on the pump energy in CoGd alloys was found by Mekonnen \textit{et~al.}~\cite{Mekonnen:2011}, which was explained by this effect. The steadily increasing behavior of $\alpha_\mathrm{eff}$ shown in Fig.~\ref{fig:freq}~\textbf{b} is well in line with the expected monotonous increase towards $L_\mathrm{comp}$. Furthermore, nonlocal damping effects due to spin-pumping either across Co/Gd interfaces or through the Co/Cu interface could contribute to the total effective damping.

In terms of the phase of the EXM, a slight variation as a function of Gd thickness was observed (not shown), which could be expected, considering the fact that EXMs are excited by a bipolar spin current pulse acting on timescales similar to those of the precessions. A similar analysis as in Ref.~\cite{Lichtenberg:2022} could be carried out, where the phase of THz standing spinwave excited by the same ps bipolar spin current pulse is investigated. However, this would go beyond the scope of this work.

\subsection{Exchange-dominated precession amplitudes}
\begin{figure}
    \centering
    \includegraphics[width=8cm]{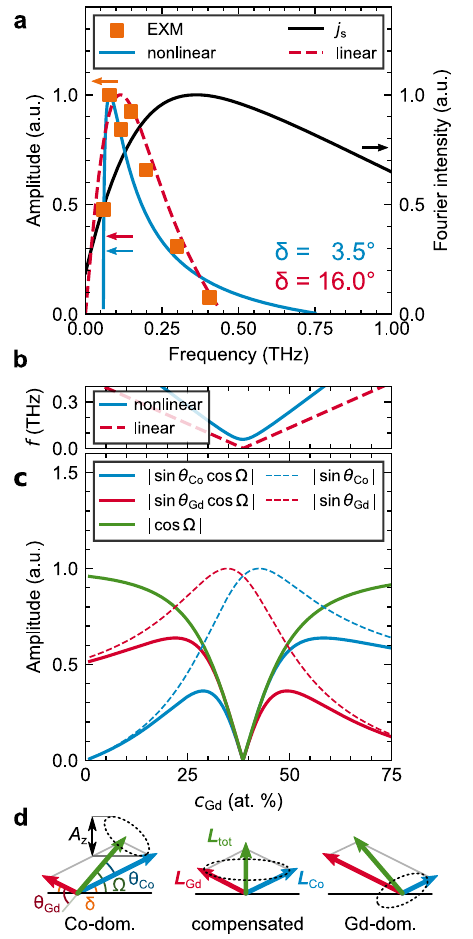}
    \caption{\textbf{a}~Amplitude of measured EXMs, fitted analytical models, and Fourier intensity of the optically induced spin current pulse as a function of frequency. The fits are based on the frequency (\textbf{b}) and amplitude (\textbf{c}) dependencies of EXMs on the alloy concentration. Designated angles are explained by the cartoon in \textbf{d}, showing deflected angular momenta of Co and Gd for different alloy concentrations.}
    \label{fig:amplitudes}
\end{figure}

Finally, we seek to explain the observed variation in EXM amplitude for different oscillation frequencies, which is shown in Figure~\ref{fig:amplitudes}~\textbf{a}, to gain both a better understanding of the system and insights into the exact excitation mechanism. A naive first guess could be the assumption that the bipolar spin current resonantly drives the precession mode. The EXM amplitude as a function of the frequency should then follow the Fourier Transform (FT) of the injection pulse. A comparison of said FT (continuous black line) to the measured data points (orange symbols) does show qualitative agreement in the sense that there is a maximum at a certain frequency and a decrease in amplitude towards zero and high frequencies. However, the maxima are located at different frequencies and the nature of the decay does not match at all. We note that the shape of the Fourier spectrum of the spin current pulse crucially depends on our assumption of the $\mathrm{d}M/\mathrm{d}t$ model, which is still a subject of ongoing discussion. Other models, such as the superdiffusive hot electron model~\cite{Battiato:2010, Battiato:2012}, assume generation of even faster optically induced spin currents, leading to stronger deviations from experimental data due to theoretical peaks at higher frequencies. Our assumption can therefore be seen as a conservative estimate for the high-frequency behaviour.

In the following, we employ a simple macrospin model that captures the ferrimagnetic nature of the absorption layer and model the OSTT as a canting of the macrospins with respect to the horizontal antiparallel state. We again consider a two sublattice, alloy-like case where the Gd concentration $c_\mathrm{Gd}$ is varied and both sublattices are initially canted by the same angle $\delta$ with respect to the film plane (shown schematically in Fig.~\ref{fig:amplitudes}~\textbf{d}). While, in practice, we are dealing with a quadlayer system, we argue that the fact that only two types of spin resonances are observed and the strong intermixing between thin layers make the alloy model a reasonable approximation. Note that the concentration $c_\mathrm{Gd}$ in the model relates to the magnetically active atomic fraction of Gd only, making direct comparisons to Gd thicknesses challenging. By using the same canting angle $\delta$ for both lattices, equal average spin transfer efficiencies for the Co and Gd sublattice are assumed. While a large fraction of spins are absorbed in the bottom Co layer, the strong intermixing with Gd contributes to a significant absorption of spin angular momentum by Gd. Furthermore, it has been shown that the spin coherence length in layered ferrimagnets is largely enhanced compared to ferromagnets~\cite{Yu:2019} leading to considerable OSTTs even deeper into the Gd layers.

Using the model defined above, we aim at understanding the anomalous dependence of the EXM amplitude as a function of frequency. Following initial canting, conservation of angular momentum dictates the subsequent resonance be around the total angular momentum vector $\mathbf{L}_\mathrm{tot}=\mathbf{L}_\mathrm{Co}+\mathbf{L}_\mathrm{Gd}$, enclosing an angle of $\Omega$ with the film plane (Fig.~\ref{fig:amplitudes}~\textbf{d}, green arrow). The vertical precession amplitude of $\mathbf{M}_\mathrm{Co}$, which is the one that is probed during the experiment, then depends on $\Omega$ and the angle $\theta_\mathrm{Co}$ that is enclosed by $\mathbf{L}_\mathrm{tot}$ and $\mathbf{M}_\mathrm{Co}$ (schematically shown in Fig.~\ref{fig:amplitudes}~\textbf{d}) according to
\begin{equation}
    A_\mathrm{z}=2|\sin \theta_\mathrm{Co} \cos\mathrm{\Omega}|.
\end{equation}
We consider two cases; in the linear approximation, the precession frequency is determined only by the degree of static angular momentum compensation of the sample with $f_\mathrm{EXM}$ given by eq.~(\ref{eq:EXM}). In other words, we assume the small angle approximation. $A_\mathrm{z}$ as a function of frequencies is fitted to the data by first converting from $f_\mathrm{EXM}$ to $c_\mathrm{Gd}$ and optimizing the free parameters $\lambda$ and $\delta$. Values for saturation magnetization and g-factors of Co and Gd are given in the Appendix. The dashed red line in Fig.~\ref{fig:amplitudes}~\textbf{a}, obtained for the linear case with $\delta=16.0^\circ$, shows good agreement. This result implies that, after the spin current pulse is fully absorbed, the magnetization vectors of Co and Gd are canted out of the film plane by this angle. In Figure~\ref{fig:amplitudes}~\textbf{b} and~\textbf{c}, the underlying frequency and amplitude dependence on the alloy concentration are shown. To understand the evolution of the amplitude, we start by considering the Co-dominated case, corresponding to low values of $c_\mathrm{Gd}$ (or $t_\mathrm{Gd}$ in the case of the experiment). Due to the strong degree of uncompensation, the oscillation frequency is high and $\mathbf{L}_\mathrm{tot}$ is very close to $\mathbf{L}_\mathrm{Co}$. As a result, $\theta_\mathrm{Co}$ is very small, leading to small $A_\mathrm{z}$. Upon increasing $c_\mathrm{Gd}$, $f_\mathrm{EXM}$ decreases and $\mathbf{L}_\mathrm{tot}$ is tilted away from $\mathbf{M}_\mathrm{Co}$, inducing larger precession amplitudes. When approaching $L_\mathrm{comp}$, an injection of OOP spins tilts $\mathbf{L}_\mathrm{tot}$ fully OOP. Therefore, the projection of the precession onto the z-axis vanishes as $\cos \Omega$ approaches zero (see Figure~\ref{fig:amplitudes}~\textbf{c}). This explains the decrease of precession amplitude in our experiments at low EXM frequencies close to $L_\mathrm{comp}$.

Since the model treated thus far is based on excited states far away from the antiparallel equilibrium state, one may argue that the commonly used small angle approximation does not hold here. To account for finite precession amplitudes, we derive the dispersion relation for arbitrary precession angles in the Appendix. The main difference to the linear approximation is that the EXM frequency remains finite for nonzero canting angles, even at the compensation point of angular momentum. The larger the precession angles are, the higher is the predicted frequency. We repeat the fitting procedure and obtain the continuous blue curve in Fig.~\ref{fig:amplitudes}~\textbf{a}. Again, the fit agrees reasonably with experimental data. For the nonlinear case, only a canting of $3.5^\circ$ is required to explain the observed trend in amplitude while the frequency is not required to cross zero (see Fig.~\ref{fig:amplitudes}~\textbf{b}). To obtain a realistic estimate for the upper bound of the canting angle, we consider the amount of angular momentum that is dissipated during ultrafast demagnetization and, based on this, calculate by how much the Co sublattice can cant upon full absorption of said angular momentum. From MOKE measurements of the injection stack, we extract a maximum demagnetization of $\sim 12.5$\%. Further, we take a thicknesses of 3.4~nm and 1.6~nm for injection and the sum of both Co layers, respectively, and use the saturation magnetizations given in the Appendix. We find a theoretical maximum of the canting angle of $7.0^\circ$. In light of this, the results from the nonlinear model seem more realistic and can explain all experimental observations.

The hitherto assumed equal spin transfer efficiency on Co and Gd requires further attention. The EXM frequency predicted by our model is actually independent of the precise ratio of $\delta_\mathrm{Co}/\delta_\mathrm{Gd}$ and thus, differences in spin transfer efficiencies. Instead, the sum $\delta_\mathrm{Co}+\delta_\mathrm{Gd}$ alone is decisive, meaning that a variety of combinations of $\delta_\mathrm{Co}$ and $\delta_\mathrm{Gd}$ can yield the same frequency. This is due to the assumed isotropy of the system. Possible anisotropy fields are much lower than exchange fields between Gd and Co and would only induce small perturbations. The z-component of the precession amplitude, on the other hand, does depend on $\delta_\mathrm{Co}/\delta_\mathrm{Co}$. We did confirm that the applicability of our model is not strictly limited to the case that $\delta_\mathrm{Co}=\delta_\mathrm{Gd}$ by fitting the data with various ratios of $\delta_\mathrm{Co}/\delta_\mathrm{Gd}$ and obtaining good results. However, it was not possible to reliably extract the best ratio from the fits.
Furthermore, different efficiencies and variations in absorption length of Co and Gd might have a minor influence on the excitation efficiency. Since dramatic changes in EXM dynamics were observed among adding less than one monolayer of Gd, we conclude that the mechanisms discussed thus far dominate.

Another interesting aspect of the dynamics at $L_\mathrm{comp}$ is that half an oscillation around $\mathbf{L}_\mathrm{tot}$ corresponds to switching the magnetization by 180$^\circ$ in the film plane. Theoretical studies on antiferromagnets~\cite{Cheng:2015, Weissenhofer:2023} came to the similar conclusion that a perpendicular spin current pulse can manipulate the order parameter of a magnetic system. Since our TR-MOKE setup is insensitive to changes in IP magnetization, we refrain from claims as to whether such switching is realistically possible in our device. We further note that distinct IP easy axes would be required to establish defined states between which the magnetization can be switched.


Finally, we will put our results into perspective with previous studies on EXMs. While higher resonance frequencies than the ones we find have been observed in insulating ferri- and antiferromagnets,~\cite{Zhang:2015, Tzschaschel:2017, Hsu:2020} those materials are extremely challenging to be implemented into devices due to their poor conductivity and the lack of conduction spin polarization. Exchange resonance modes in a similar frequency range ($\sim0.4$~THz) have not been reported in metallic systems at room temperature to the best of our knowledge. Comparable Ru-based synthetic antiferromagnetic oscillators only reach frequencies in the $\sim 20$~GHz range at zero field~\cite{Waring:2020} which is owed to the RKKY coupling being much weaker than the RE-TM exchange. The sample design used in our study offers not only a high conductivity in general but also ferromagnetic interfacial layers with high conduction spin polarization. Consequently, our layer structure could easily be integrated into electronic applications, as magnetoresistive effects can be used to probe the precession state on-chip. Furthermore, our results deepen understanding on intricacies in the fascinating platform of Co/Gd multilayers that are highly relevant for other types of spintronic applications such as magnetic racetrack memory devices.

It is worth noting that the coherent excitation of THz dynamics by optically induced spin currents is not limited to the sample design used in this work. Instead, it can easily be adapted to study high-frequency modes in other types of ferrimagnets or even antiferromagnets.

\section{Conclusion}
We have demonstrated excitation of ferromagnetic resonance and exchange resonance modes in IP synthetic ferrimagnetic Co/Gd/Co/Gd multilayers using OSTTs generated in a neighboring perpendicular magnetic layer. Optically induced spin currents were proven to be an excellent tool to excite and study those modes. FMR oscillations in the 10~GHz~range and EXM modes with frequencies up to 0.4~THz were observed at room temperature. Varying the thickness of the Gd layers was investigated, enabling an easy path for manipulating the spin resonance spectrum over orders of magnitude by tuning the total angular momentum in the system. The dependence of the frequency of the exchange mode gives unique insight into the excitation mechanism and possible nonlinear dynamics close to compensation. Our findings open up new pathways for the development of ferrimagnetic THz spintronic devices and exploring their high-frequency response.

\section*{Acknowledgments}
This project has received funding from the European Union’s Horizon 2020 research and innovation programme under the Marie Skłodowska-Curie grant agreement No 861300.

\appendix*

\section{Nonlinear exchange resonance dynamics}
The equations of motion predicting the time evolution of two antiferromagnetically coupled magnetic moments $M_\mathrm{Co,Gd}$ in the absence of external fields, magnetic anisotropy, and Gilbert damping are given by the following coupled Landau-Lifshitz (LL) equations:
\begin{equation}\label{eq:LLG}
    \begin{aligned}
        \frac{\mathrm{d}\mathbf{M}_\mathrm{Co}}{\mathrm{d}t} &= -\mu_0\gamma_\mathrm{Co}\mathbf{M}_\mathrm{Co}\times\lambda\mathbf{M}_\mathrm{Gd},\\
        \frac{\mathrm{d}\mathbf{M}_\mathrm{Gd}}{\mathrm{d}t} &= -\mu_0\gamma_\mathrm{Gd}\mathbf{M}_\mathrm{Gd}\times\lambda\mathbf{M}_\mathrm{Co}.
    \end{aligned}
\end{equation}

\begin{figure}[htbp]
    \centering
    \includegraphics[width=7cm]{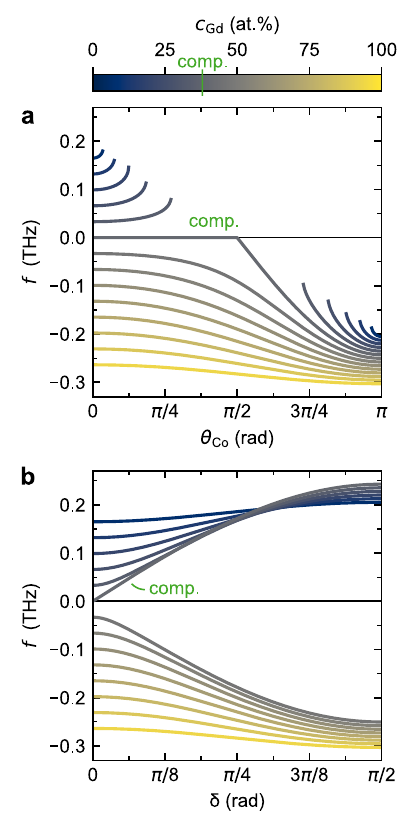}
    \caption{Exchange resonance frequency \textbf{a}~as a function of $\theta_\mathrm{Co}$ according to eq.~(\ref{eq:EXMnl}) and \textbf{b}~as a function of $\delta$ as given by eq.~(\ref{eq:EXMdelta}) for different alloy concentrations. A change in sign of the frequency implies a reversal of the mode's handedness.}
    \label{fig:ftheo}
\end{figure}

Upon canting the magnetic moments by an angle $\delta$ with respect to the film plane, as explained in the main text, conservation of angular momentum requires the consequent precessions to revolve around $\mathbf{L}_\mathrm{tot}$, enclosing an angle of
\begin{equation}
    \Omega=\mathrm{arctan}\frac{\sin(\delta)\left[\frac{M_\mathrm{Co}}{\gamma_\mathrm{Co}} + \frac{M_\mathrm{Gd}}{\gamma_\mathrm{Gd}}\right]}{\cos(\delta)\left[\frac{M_\mathrm{Co}}{\gamma_\mathrm{Co}} - \frac{M_\mathrm{Gd}}{\gamma_\mathrm{Gd}}\right]}
\end{equation}
with the film plane. The angles of $\mathbf{M}_\mathrm{Co,Gd}$ with respect to $\mathbf{L}_\mathrm{tot}$ are given by $\theta_\mathrm{Co}=\Omega-\delta$ and $\theta_\mathrm{Gd}=\Omega+\delta$ (graphically shown in Fig.~\ref{fig:amplitudes}~\textbf{d}). As the angular momentum perpendicular to the rotation axis has to vanish, the following relation between the two angles holds:
\begin{equation}\label{eq:CAM}
    \frac{M_\mathrm{Co}}{\gamma_\mathrm{Co}}\sin\theta_\mathrm{Co}=\frac{M_\mathrm{Gd}}{\gamma_\mathrm{Gd}}\sin\theta_\mathrm{Gd}.
\end{equation}
Equation~(\ref{eq:LLG}) may now be solved for the precession frequency assuming an oscillatory solution and identifying angles between $\mathbf{H}^\mathrm{ex}_\mathrm{Gd,Co}$ and $\mathbf{M}_\mathrm{Co,Gd}$. The LL equations then simplify to
\begin{align}
        2\pi f M_\mathrm{Co} \sin \theta_\mathrm{Co} &= \mu_0\gamma_\mathrm{Co}\lambda M_\mathrm{Co}M_\mathrm{Gd}\sin (\theta_\mathrm{Gd}-\theta_\mathrm{Co}),\\
        2\pi f M_\mathrm{Gd} \sin \theta_\mathrm{Gd}&= -\mu_0\gamma_\mathrm{Gd}\lambda M_\mathrm{Gd}M_\mathrm{Co}\sin (\theta_\mathrm{Co}-\theta_\mathrm{Gd}).
\end{align}
The constraint given by eq.~(\ref{eq:CAM}) makes the two solutions equivalent. A simple rearrangement of the solution for the Co sublattice yields
\begin{align}
    f&=\frac{\mu_0\gamma_\mathrm{Co}\lambda M_\mathrm{Gd}\sin (\theta_\mathrm{Gd}-\theta_\mathrm{Co})}{2\pi\sin\theta_\mathrm{Co}}\label{eq:EXMnl}\\
    &=\frac{\mu_0\gamma_\mathrm{Co}\lambda M_\mathrm{Gd}\sin (2\delta)}{2\pi\sin\theta_\mathrm{Co}}.\label{eq:EXMdelta}
\end{align}

The former equation as a function of $\theta_\mathrm{Co}$ and the latter one as a function of the canting angle $\delta$ are plotted in Fig.~\ref{fig:ftheo}~\textbf{a} and \textbf{b}, respectively for a variety of alloy compositions and an arbitrarily chosen exchange constant of $\lambda=2$. Both solutions reproduce the change in sign of the frequency across $L_\mathrm{comp}$, implying a reversal of handedness. Increasing $\theta_\mathrm{Co}$ and $\delta$ both leads to an increase of $|f|$ with respect to the value obtained at $\theta_\mathrm{Co}=\delta=0$. In the Co-dominated region, dispersion curves in Fig.~\ref{fig:ftheo}~\textbf{a} show discontinuities whenever eq.~(\ref{eq:CAM}) has no real value solution for $\theta_\mathrm{Gd}$. Figure~\ref{fig:ftheo}~\textbf{b} is only zero when $L$ is compensated and $\delta=0$. Therefore, any canting away from the antiparallel ground state in a ferrimagnet will result in an exchange-driven spin resonance. One can easily show that eq.~(\ref{eq:EXMnl}) simplifies to eq.~(\ref{eq:EXM}) from the main text in the small angle approximation by linearizing trigonometric functions and substituting eq.~(\ref{eq:CAM}) for $\theta_\mathrm{Gd}$.

\begin{table}[htbp]
\caption{\label{tab:parameters}%
Saturation magnetizations and $g$-factors of Co and Gd used in the macrospin model.}
\begin{ruledtabular}
\begin{tabular}{ll}
    $M_\mathrm{Co}$ & 1.44~MA/m~\cite{Coey:2010}\\
    $M_\mathrm{Gd}$ & 2.06~MA/m~\cite{OHandley}\\
    $M_\mathrm{Co/Ni}$ & 0.66~MA/m~\cite{Lalieu:2017b}\\
    $g_\mathrm{Co}$ & 2.17~\cite{Coey:2010}\\
    $g_\mathrm{Gd}$ & 1.95~\cite{Coey:2010}
\end{tabular}
\label{tab:constants}
\end{ruledtabular}
\end{table}

\bibliography{ref}

\end{document}